\documentclass[11pt,a4paper]{article}
\usepackage{jcappub}

\def\be{\begin{equation}}
\def\ee{\end{equation}}
\def\beq{\begin{equation}}
\def\eeq{\end{equation}}
\def\bea{\begin{eqnarray}}
\def\eea{\end{eqnarray}}
\def\bml{\begin{subequations}}
\def\blea{\bml\begin{eqnarray}}
\def\elea{\end{eqnarray}\end{subequations}}

\begin{document}

\title{The Continuum of Discrete Trajectories in Eternal Inflation}

\author{Vitaly Vanchurin}

\emailAdd{vvanchur@umn.edu}

\date{\today}

\affiliation{Department of Physics, University of Minnesota, Duluth, Minnesota, 55812}

\abstract{

We discuss eternal inflation in context of classical probability spaces defined by a triplet: sample space, $\sigma$-algebra and probability measure.  We show that the measure problem is caused by the countable additivity axiom applied to the maximal $\sigma$-algebra of countably infinite sample spaces. This is a serious problem if the bulk space-time is treated as a sample space which is thought to be effectively countably infinite due to local quantum uncertainties. However, in semiclassical description of eternal inflation the physical space expands exponentially which makes the sample space of infinite trajectories uncountable and the (future) boundary space effectively continuous. Then the measure problem can be solved by defining a probability measure on the continuum of trajectories or holographically on the future boundary. We argue that the probability measure which is invariant under the symmetries of the tree-like structure of eternal inflation can be generated from the Lebesgue measure on unit interval. According to Vitali theorem the Lebesgue measure leaves some sets without a measure which means that there are certain probabilistic questions in eternal inflation that cannot be answered.   

}

\maketitle

\section{Introduction}

One of the most (if not the most) important unresolved problem in modern cosmology is the problem of assigning probabilities to cosmological observations. In context of the big bang theory the problem is known as the problem of initial conditions \cite{InitialConditions,InitialConditions2,InitialConditions3,InitialConditions4} and in context of eternal inflation the problem is known as the measure problem \cite{EternalInflation,EternalInflation2,EternalInflation3,EternalInflation4}. In this paper we will concentrate on the measure problem for semiclassical models of eternal inflation where the underlying geometry is treated completely classically. For most of the discussion we will remain within the axiomatic framework of the classical probability spaces, but possible departures form the realm of classical probabilities will also be discussed. 

The classical probability spaces consist of three essential ingredients: a sample space of elementary events, a $\sigma$-algebra of events, and a probability measure \cite{ClassicalProbabilities}. In the models of eternal inflation the sample space is usually a single (but infinite) realization of the classical space-time, and the remaining task (or the everlasting measure problem \cite{MeasureProblem}) is to define a $\sigma$-algebra and a probability measure which would not suffer from the many problems and paradoxes including  entropy problem \cite{Entropy,Entropy2,Entropy3,Entropy4,Entropy5,Entropy6}, youngness paradox \cite{Youngness,Youngness2,Youngness3}, Boltzmann brains problem \cite{Boltzmann,Boltzmann2,Boltzmann3}, Guth-Vanchurin paradox \cite{GuthVanchurin,GuthVanchurin2,GuthVanchurin3,GuthVanchurin4}, etc. The most obvious choice is to construct a maximal $\sigma$-algebra which is a power set of the sample space, but that does not work for (countably) infinite sample spaces as will be discussed in the next section. This indeed would be a serious (or ill-defined) mathematical problem if our sample space was a set of disconnected elementary events with no additional structure, but in eternal inflation (or at least in a semiclassical model of eternal inflation) the events live in the same space-time and the isometries of the space-time must be respected in constructing probability spaces. In this paper we will use the tree-like structure of eternal inflation to define a sample space of trajectories, to construct a $\sigma$-algebra on the space of trajectories and to derive a probability measure which is invariant under the symmetry transformations. The measure will leave some of the sets non-measurable, which is the price we pay for defining probabilities on a continuum of infinite trajectories.

The paper is organized as follows. In the next section we define classical probability spaces, identify the measure problem for countably infinite sets, and discuss possible generalizations of the classical probabilities. In Sec. \ref{Tree} we study the structure of binary trees, calculate cardinalities of different sets, and construct Lebesgue measure on the tree. In the Sec. \ref{Inflation} we apply the framework of classical probabilities to the tree-like structure of eternal inflation by defining a non-maximal $\sigma$-algebra and a uniform probability measure. The main results are summarized in Sec. \ref{Summary}.

\subsection{Probability Spaces}\label{Spaces}

The probability spaces are usually (but not always) defined by a triplet $(\Omega, {\cal F}, P)$, where the sample space, $\Omega$, is a set elementary events, the ${\cal F}$ is a set of subsets of $\Omega$, and the probability measure, $P$, is a map from $\cal F$ to the unit interval $ [0,1]$.  An important requirement on the set of subsets ${\cal F}$ that it must be a $\sigma$-algebra, i.e. non-empty (${\cal F}\neq \emptyset$) and closed under complementation ($e \in {\cal F} \Rightarrow \Omega \setminus e \in {\cal F}$) and countable union ($e_i \in {\cal F} \Rightarrow \cup_{i=1}^\infty e_i \in {\cal F}$). One example of the $\sigma$-algebra ${\cal F}$ over $\Omega$ is a set of all subsets (also known as a power set) usually denoted by $2^\Omega$. 

For the triplet $(\Omega, {\cal F}, P)$ to be regarded as a classical probability space it must also satisfy Kolmogorov's probability axioms:\\
{\it Positivity}: Probability of any one event is a non-negative real number:
\be
P(e) \ge 0 \;\; \forall \;\; e \in {\cal F};
\label{CPositivity}
\ee
{\it Unitarity}: Probability of all of the events is unity:
\be
P(\Omega)=1;
\label{CUnitarity}
\ee
{\it Additivity}: Probability of disjoint events is additive:
\be
P(\cup_{i=1}^\infty e_i) = \sum_{i=1}^\infty P(e_i) \;\;\;\;\;\forall \;\;e_i \in {\cal F}\;\;\;\; \text{and} \;\;\;\; i\neq j \;\; \Rightarrow \;\; e_i \cap e_j =  \emptyset .
\label{CAdditivity}
\ee
For example, the classical probability space of a fair coin or of a fair random bit is described by 
$\Omega=\{0, 1\}$, 
 ${\cal F} = \{\emptyset, \{ 0 \}, \{1\}, \{0, 1\}\}$ and 
 $P(\emptyset) =0, P(\{0\})=1/2, P(\{1\})=1/2, P(\{0,1\})=1$ and it is a straightforward exercise to check that all three probability axioms are satisfied. Here ${\cal F}$ is the maximal $\sigma$-algebra which is the power set of the sample space, but this need not be the case in general.

Given a countably infinite sample space (as is often assumed in context of eternal inflation, but will be disputed later in the paper),  the next step could have been the construction of a probability measure $P:{\cal F} \rightarrow [0,1]$ with $\sigma$-algebra ${\cal F}$ defined as a power set $2^\Omega$, but we immediately  encounter a problem. The problem is that if we are to assign equal and finite probabilities to a countable infinity of elementary events, i.e. 
\be
P(\{ \omega \}) = \text{const} > 0 \;\;\; \forall \omega \in \Omega
\label{FiniteProb}
\ee
then the (countable) additivity axiom \eqref{CAdditivity} would imply $P(\Omega) = \infty$ in conflict with the axiom of unitarity \eqref{CUnitarity}. On the other hand if we assign zero probability to all of the elementary events, i.e. 
\be
P(\{ \omega \}) = 0 \;\;\; \forall \omega \in \Omega
\ee
then the (countable) additivity axiom \eqref{CAdditivity} would imply $P(\Omega) = 0$ which is once again in a conflict with \eqref{CUnitarity}.  The problem could be avoided if the additivity axiom \eqref{CAdditivity} is weakened or replaced \cite{Finetti}, for example,   with\\
{\it Finite Additivity}: 
\be
P(e_1\cup e_2) = P(e_1) + P(e_2) \;\;\;\;\;\forall \;\;e_1, e_2\in {\cal F}\;\;\;\; \text{and} \;\;\;\; e_1 \cap e_2 =  \emptyset.
\label{Finite}
\ee
or with\\
{\it Independent Additivity}:
\be
P(e_1 \cup e_2) = P(e_1) + P(e_2) - P(e_1) P(e_2) \;\;\;\;\;\forall \;\;e_1,e_2 \in {\cal F}
\label{Independent}
\ee
Then if we assign zero probability to all of the elementary events \eqref{FiniteProb}, the finite additivity \eqref{Finite} or independent additivity \eqref{Independent} conditions would imply that the size of all finite sets of the elementary events or finite subsets of $\Omega$ is exactly zero. Moreover, the unitarity axiom \eqref{CUnitarity} and the independent additivity \eqref{Independent} combined imply that all of the infinite subsets must have probability one, i.e.
\be
P( e) = \begin{cases} 0 \;\; \text{if} \; |e| < \infty \\  1 \;\; \text{if} \; |e| = \infty  \end{cases}.
\ee
Also note that when compared with the standard addition rule 
\be
P(e_1 \cup e_2) = P(e_1) + P(e_2) - P(e_1 \cap e_2)\;\;\;\;\;\forall \;\;e_1,e_2 \in {\cal F}
\ee
the independent additivity axiom \eqref{Independent} suggests that the probabilities of the intersection of events must be calculated not as the intersections of subsets, but as if the events are statistically independent.\footnote{For the sake of completeness note that the problem of defining probabilities on a countable set of elements can also be addressed in the context of the non-Archimedean numbers such as hyperreals (or nonstandard) reals \cite{Nelson}. In addition to real numbers the hyperreal numbers includes infinitesimals which are strictly greater than zero, but smaller than $1/n$ for any $n\in\mathbb{N}$. Then one can assign equal infinitesimal probabilities to all of the elementary events and at the same time demand that the additivity axiom holds in such a way that the probability of a countable union of elementary events is a finite number. The study of such non-Archimedean probability spaces in the context of eternal inflation is certainly interesting avenue to explore, but is beyond the scope of this paper.}

Although the possible modifications of the classical probabilities deserve to be explored further, in this paper we will restrict our attention on the probabilities defined by the Kolmogorov axioms \eqref{CPositivity}, \eqref{CUnitarity} and \eqref{CAdditivity}. Evidently, the most problematic axiom in defining uniform probability measures on a countable set of elementary events is the additivity axiom due to the countable sums and unions which appear in \eqref{CAdditivity}. This suggests that within the framework of classical probabilities the problem can be avoided if the sample space contains either a finite or an uncountable number of elements. Since the former case is trivial, let us study the latter case when the sample space forms a continuum of real numbers between $0$ and $1$. There exists a well known uniform measure on the unit interval known as Lebesgue measure, $P_L$, which assigns zero probability to all elementary events (i.e. real numbers on the unit interval) and as a consequence of the additivity axiom \eqref{CAdditivity} to all countable collections of elementary events. For example the measure of all rational numbers on the unit interval is exactly zero, $P_L([0,1]\cap \mathbb{Q})=0$. The reason it is no longer in a conflict with unitarity axiom \eqref{CUnitarity} is that according to the Lebesgue measure only (but not all) subsets with uncountable number of elements, such as the entire sample space, can have non-zero measure, $P_L([0,1])=1$. However,  the Lebesgue $\sigma$-algebra is not a maximal $\sigma$-algebra, as there are sets of real numbers (e.g. Vitali sets) that are not in the Lebesgue $\sigma$-algebra and thus are not measurable with respect to the Lebesgue measure \cite{Vitali}.  It was also shown by Hausdorff that such non-measurable sets must exist for any measure on real numbers which is invariant under symmetry transformations such as translations.

\section{Binary Tree}\label{Tree}

Let us highlight the two most important points that we discussed in the previous section. The first point is that the measure problem is due to the countable additivity axiom \eqref{CAdditivity} applied to the maximal $\sigma$-algebra of countably infinite sample spaces in eternal inflation. Thus to avoid the problem one must either construct a non-maximal $\sigma$-algebra or to define a different sample space with either a finite or uncountably many elementary events.  The second point is that the probability measure must respect the isometries of the sample space even if the construction leaves certain sets non-measurable. In what follows we will define an uncountable sample space, a non-maximal $\sigma$-algebra and construct an invariant probability measure on an infinite binary tree whose internal structure resembles the most essential isometries in eternal inflation \cite{Tree}. 

\begin{figure}[]
\includegraphics[width=0.75\textwidth]{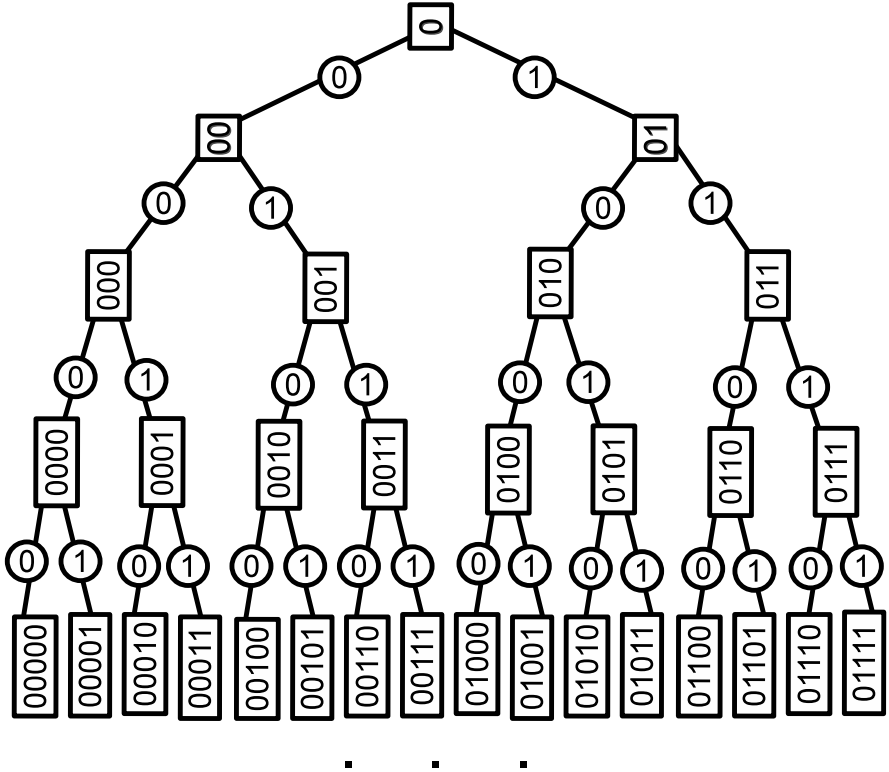}
\caption{Binary Tree. \label{plot}}
\end{figure}

To discuss classical probability spaces on a sample space of infinite rooted tree it will be convenient to label all of the nodes and edges of the tree. We will describe the procedure for a binary tree,  $T_2$,  but it can be easily generalized for an arbitrary $k$-ary tree,  $T_k$. First we label each edge with either $0$ or $1$ in such a way that each node has a single $0$ edge (or ``left edge'') and a single $1$ edge  (or ``right edge'') connected to its children nodes. Then we label the root  node of the tree with $0$ (it could have been $1$ without loss of generality) and all other nodes inductively with the same label as the label of its parent node concatenated with either $0$ or $1$ depending on whether the connecting edge is labeled with $0$ or $1$. (See Fig. \ref{plot}) 

 For example, the first three generations (or levels or depth) of nodes will have the following labels: $0, 00, 01, 000, 001, 010, 011$. What is different now, compared to the discussion of the previous section, is that the nodes are no longer equivalent (or symmetric) to each other. Although the large symmetry is broken by the structure of the graph there is a residual symmetry that we must take into account when classical probability spaces are defined on the tree.  In particular, we will demand that the permutation of edges of any node (or the exchange of  $0$ and $1$ edges) would leave the probability measure invariant. 

The cardinality of the set of nodes in the infinite rooted binary tree (as well as $k$-ary tree) is that of a countable infinity,
\be
|T_2| = \aleph_0.
\ee
This can be shown explicitly by constructing a one-to-one map between labels of all nodes, $x$,  and the set of natural numbers,
\be
f(x) =  2^{l(x)-1}+n(x)
\ee
where $l(x)$ is the length of the word $x$ and $n(x)$ is the numerical value of the word $x$ written in binary code. Then the cardinality of the power set of the set of nodes is a continuum, 
\be
\left | 2^{T_2}\right | = 2^{\aleph_0}.
\ee
Moreover, the cardinality of the set of infinite trajectories (or paths of infinite depth) in the binary tree is also a continuum. This could not be anticipated from considering finite portions of the tree as is usually done in the context of eternal inflation: for any finite cut-off at depth $l$ the number of nodes, $2^l-1$, is greater or equal than the number of trajectories, $2^{l-1}$, which start at the root and terminate at the cut-off. In the limit of large $l$ the number of nodes is twice as much as the number of trajectories and one could have (naively) expected that the number of infinite trajectories cannot be larger than the number of nodes, but this turns out not to be the case.

To prove that the set of infinite trajectories is much bigger than the set of nodes we can use Cantor's diagonal argument.  The arguments starts with an assumption that there is a one-to-one map $S$ from natural numbers $k$ to all of the infinite trajectories $S(k)$ which in our case are labeled by infinite words. Let us also denote the value of the $k$'s digit (or letter) by $x^k$, the concatenation operation (i.e. $[i, n, f, l, a, t, i, o, n] = inflation$) and the logical `not' function (i.e. $\lnot 0 = 1$ and $\lnot 1 = 0$). Then we can construct a word 
\be
X= [0, \lnot S(1)^2, \lnot S(2)^3, \lnot S(3)^4 ...]
\ee
which was not mapped to by the map $S$ and thus a contradiction is reached. This means that the cardinality of the set of infinite trajectories is bigger than $\aleph_0$ and in fact is the same as the cardinality of a continuum, $2^{\aleph_0}$. Then we could try to generate the smallest $\sigma$-algebra of the sample space of nodes which contains all of the sets of infinite trajectories, but unfortunately such $\sigma$-algebra is maximal which comes with the baggage of problems discussed above. However, the fact that the set of infinite trajectories forms a continuum of configurations suggests that we might what to use this set as a sample space instead of the set of nodes. (We denote both sets with $T_2$, but the distinction would be clear from the context.) 

Given that the cardinality of the set of infinite trajectories in $T_2$ is the same as the cardinality of the set of real numbers on unit interval $[0,1]$ we should be able to construct a one-to-one map between the two sets.  For example, one could define a map 
\be
g(x) \equiv n([0, ., x^2, x^3, ...]), 
\label{Map}
\ee
but this map is not one-to-one since there are different trajectories that are mapped to the same real number, e.g. $0111...$ and $1000...$ are mapped to $g(0111...)=g(1000...)=0.1$. Fortunately there is only a countable number of such reals 
\be
A=\left \{0, 1, \frac{1}{2} , \frac{1}{4}, \frac{3}{4}, \frac{1}{8}, \frac{3}{8}, \frac{5}{8}, \frac{7}{8}, ... \right \}
\ee
and only a countable number of such trajectories
\be
B=\{00000..., 01111..., 01000..., 00111..., 00100..., 00011..., ...\}.
\ee
and, thus,  we can modify $g(x)$ to make it one-to-one
\be
\tilde{g}(x)\equiv \begin{cases} n([0, ., x^2, x^3, ...]) &\;\;\text{if}\;\; x \notin B\\
A_n &\;\;\text{if}\;\; \exists\;n\in \mathbb{N} \mid  x=B_n.
\end{cases}
\ee
However, the fact that the two maps $\tilde{g}$ and $g$ differ only in the way a small (countable) number of elements is mapped we can use either map to define a probability measure on  the tree.

For example, we can use the map $g: T_2 \rightarrow [0,1]$ to generate a $\sigma$-algebra 
 \be
{\cal L}(T_2) = \{  g^{-1}(e) \mid e \in {\cal L}([0,1])  \}
\ee
(where $g^{-1}(e)$ is the preimage of $e$) and a probability measure on infinite trajectories,
\be
P_{L(T_2)}(e) = P_L(g(e)) \;\; \forall e \in {\cal L}(T_2)
\ee
where ${\cal L}([0,1])$ and $P_L$ is respectively the Lebesgue $\sigma$-algebra and the Lebesgue measure on unit interval.
We will refer to this measure on the tree as Lebesgue measure since it is generated from the Lebesgue measure on unit interval, but it remains to be checked that the measure is invariant under the symmetry transformations of the tree. The symmetry group on the graph of the binary tree is quite large and is generated by operations of swapping $0$ and $1$ labels on edges from parent node to its children. However, due to the symmetry of our construction all of these transformations must leave the Lebesgue measure on the binary tree invariant similarly to how translations leave the Lebesgue measure on the real line invariant. Then according to Vitali theorem the Lebesgue $\sigma$-algebra is not maximal and thus there are sets of infinite trajectories that are also non-measurable (not to confuse with sets of measure zero). The existence of non-measurable sets, also known as Vitali sets, relies on the axiom of choice and is also known to lead to the famous Banach-Tarski paradox of how a single ball can be decomposed and resembled into two balls identical to the original \cite{Banach}.

Let us stop for a second to emphasize this important result. In the first section when measures were constructed on infinite sets of elementary events with no additional structure the only reasonable $\sigma$-algebra of events seems to be the maximal $\sigma$-algebra. Then the construction of a uniform probability measure was not possible within the frameworks of classical probabilities unless some of the axioms were weakened or replaced. Now that we have the additional tree-like structure it seems possible to define a Lebesgue measure on the sample space of infinite trajectories, but the corresponding Lebesgue  $\sigma$-algebra is not maximal which leads to non-measurability of certain events.

\section{Eternal Inflation}\label{Inflation}

We are now ready to define a sample space of the elementary events, on the space-time generated by eternal inflation. Since the elementary events will be associated with possible observations there are quantum and gravitational limitations that should be taken into account. The Hilbert spaces of a single harmonic oscillator can be described with a countable number of orthonormal basis and thus is separable.  If we take a finite collection of such oscillators (coupled or uncoupled) the dimensionality of the Hilbert space remains only countably infinite. However, if we consider an infinite lattice of countably many harmonic oscillators in each lattice point (as in field theories) the dimensionality becomes uncountably infinity. The same thing happens for a countable collection of quantum bits whose Hilbert space has an uncountably infinite dimensionality. This implies that the space of all possible outcomes of measurements or the sample space is uncountable and is the same as the cardinality of the power set of a countable set (i.e. a continuum). 

This is what one would generically expect from a quantum theory, but there are certain limitations which come from gravity. First of all it is believed that the countable dimensionality of the Hilbert space can be reduced to a finite number once the gravitational effects are taken into account. The way it usually works is that certain states have energies (and thus masses) that would have collapsed into a black-hole and therefore must be excluded from counting. This might reduce the number of states within a finite volume to a finite number, but that does not reduce the uncountable-dimensionality of the Hilbert space in infinite space-time and thus the uncountable size of the corresponding sample space. There is also an additional gravitational effect which limits the information that can in principle be measured by a local observer. This does not apply to Minkwoski space-time, but it does apply to the space-time we study here. In eternal inflation the local degrees of freedom seem to constantly fall out of causal contact with a local observer due to exponential stretching of space in eternal inflation. One attitude is that these degrees of freedom do not disappear from the Hilbert space of the observer, but become encoded in the de Sitter radiation. In this view the dimensionality of the Hilbert space of local observers remains always finite which is all that we need to define a good sample space without having to deal with the oddities of the countable additivity axiom discussed above. Another attitude (that we will assume here) is that the new degrees of freedom are constantly stretched out from under the Planck scale and thus must be included in the Hilbert space of a local observer. Then the dimensionality of the Hilbert space and the cardinality of the sample space of local observations remains uncountably infinite and the measure problem can be reduced to the problem of defining a measure on a continuum of sequences of observations (or what we call infinite trajectories) which is invariant under whatever symmetries the eternal inflation may possess.

For example, it is expected that eternally inflating space-time resembles the tree-like structure similarly to what was discussed in the previous section. (See Ref. \cite{Tree} for the discussion of the tree-like structure of eternal inflation generated from either the scale-factor  \cite{ScaleFactor} or light-cone time  \cite{LightconeTime} coordinates using the square-bubble approximation.) Although we have only studied the binary tree, $T_2$, the analysis can be easily generalized to an arbitrary $k$-ary tree, $\Omega=T_k$, given that $k$ is finite and the same for all node. For more general models of eternal inflation with terminal vacua more general trees with perhaps a variable number of edges might be required, but it is expected that the generalization should not be too difficult. Moreover, the presence of terminal vacua would require specification of measures on fractals in which case the Lebesgue measure would have to be replaced with something like Sinai-Ruelle-Bowen measure \cite{Trajectories6}. We will intentionally avoid all these complication and will assume that eternal inflation (or at least a toy model of eternal inflation) can be reduced to study of a rooted $k$-ary tree with local measurements corresponding to nodes of the tree and edges corresponding to the exponential growth. Then our sample space of elementary events is a set of infinite trajectories (or paths) originating from the root and going all the way to the future boundary.\footnote{A possibility of defining holographic measures on a boundary is not entirely new and was considered in a number of recent publications. See for example Refs. \cite{Holographic,Holographic2,Holographic3,Holographic4,Holographic5}. Likewise the cosmological measures on infinite trajectories were considered in Refs. \cite{Trajectories6,Trajectories,Trajectories2,Trajectories3,Trajectories4,Trajectories5}.}

Then to define a probability space $(T_k, {\cal L}(T_k), P_{L(T_k)})$ on the $k$-ary tree we first map the infinite trajectories $x \in T_k$ of the tree to the unit interval using \eqref{Map} where now the digits can have a bigger range 
\be
x^n \in \{0, 1, ...,k-1\} \;\;\; \forall n \in \mathbb{N}
\ee
and then use the Lebesgue $\sigma$-algebra and Lebesgue measure to generate a $\sigma$-algebra and to define a probability measure on the tree
\bea
{\cal L}(T_k) = \{  g^{-1}(e) \mid e \in {\cal L}([0,1])  \} \\
P_{L(T_k)}(e) = P_L(g(e)) \;\; \forall e \in {\cal L}(T_k).
\eea
Similarly to the case with binary tree the Lebesgue measure on $k$-ary tree satisfies all of the symmetries of permutations of edges at every node, but it once again leaves some sets (i.e. Vitali sets) without a measure. 

Once the classical probability space $(T_k, {\cal L}(T_k), P_{L(T_k)})$  is defined we can define classical random variables (or observables) as real-valued functions on the sample space
\be
{\cal O}: T_k \rightarrow \mathbb{R}
\ee
that are measurable with respect to the $\sigma$-algebra  ${\cal L}(T_k)$ and whose statistical moments are calculated by
\be
E[{\cal O}^n] =  \int_{T_k} {\cal O}^n(\omega) dP_{L(T_k)}(\omega) \;\;\;\; \forall\; n\in\mathbb{N}
\ee
Although we are not going to discuss the observational prediction of the Lebesgue measure in details it is expected that its phenomenology will be similar to phenomenology of the scale-factor measure \cite{ScaleFactor,ScaleFactor2} or light-cone measure \cite{LightconeTime, LightconeTime2}, due to the similarities of these measures and the tree-like structure discussed in \cite{Tree}, with some important differences. First of all, the probability of all finite trajectories, whose cardinality can be shown to be that of a countable infinity, should be exactly zero. Secondly, as was already noted, the non-measurability of the Vitali sets should put constraints to what can in principle be observed. And finally, since the Lebesgue measure is not defined using cut-off prescriptions we expect that the many paradoxes of the cut-off measures may be avoided.

\section{Summary}\label{Summary}

We conclude with a summary of the main results:

{\it 1)  Additivity Axiom: } The classical probability spaces are defined by a triplet: sample space, $\sigma$-algebra and probability measure, which must satisfy the positivity \eqref{CPositivity}, unitarity \eqref{CUnitarity} and additivity \eqref{CAdditivity} axioms. In context of eternal inflation the measure problem is caused by the additivity axiom applied to the maximal $\sigma$-algebra of countably infinite sample spaces. There are however other frameworks such as non-countably-additive \cite{Finetti} or non-Archimedian \cite{Nelson} where the problem can be avoided. 

{\it 2) Continuum of Trajectories:} The quantum uncertainties make the bulk space-time effectively discrete and the corresponding sample space of local observations effectively countable, but the exponential expansion makes the future boundary space effectively continuous and the corresponding sample space of infinite trajectories effectively uncountable. Then the measure problem can be solved by defining a $\sigma$-algebra and a probability measure on the continuum of infinite trajectories  or holographically on the future boundary.

{\it 3) Lebesgue Measure:} $\sigma$-algebra and  probability measure on a continuum of infinite trajectories can be constructed by demanding that the measure is invariant under the symmetries of eternally inflating space-time. We use the symmetries of the tree-like structure of eternal inflation to derive a probability measure which is generated from the Lebesgue measure on unit interval.

{\it 4) Non-measurable Sets:} According to Vitali theorem there exist sets (e.g. Vitali sets) which are non-measurable with respect to the Lebesgue measure. This means that there are certain probabilistic questions in eternal inflation that cannot be answered. 

{\it Acknowledgments.} The author is grateful to Alan Guth, Mahdiyar Noorbala, Ken Olum and Alex Vilenkin for very useful discussions and comments on the manuscript. The work was supported in part by Templeton Foundation and Foundational Questions Institute (FQXi).

\end{document}